\documentclass[5p]{elsarticle}

\usepackage{amssymb}
\usepackage{amsfonts}
\usepackage{amstext}
\usepackage{graphicx}
\usepackage{amscd}
\usepackage{amsmath}

\newcommand{\ihbar}{\imath \hbar}
\newcommand{\tr}{\mathrm{tr}}

\begin{document}

\begin{frontmatter}

\title{Quantum chimera states}

\author[uti]{David Viennot}
\author[uti]{Lucile Aubourg}
\address[uti]{Institut UTINAM (CNRS UMR 6213, Universit\'e de Bourgogne-Franche-Comt\'e, Observatoire de Besan\c con), 41bis Avenue de l'Observatoire, BP1615, 25010 Besan\c con cedex, France.}

\begin{abstract}
 We study a theoretical model of closed quasi-hermitian chain of spins which exhibits quantum analogues of chimera states, i.e. long life classical states for which a part of an oscillator chain presents an ordered dynamics whereas another part presents a disordered dynamics. For the quantum analogue, the chimera behavior deals with the entanglement between the spins of the chain. We discuss the entanglement properties, quantum chaos, quantum disorder and semi-classical similarity of our quantum chimera system. The quantum chimera concept is novel and induces new perspectives concerning the entanglement of multipartite systems.
\end{abstract}


\end{frontmatter}

\section{Introduction}
\begin{figure}
\begin{center}
\includegraphics[width=8cm]{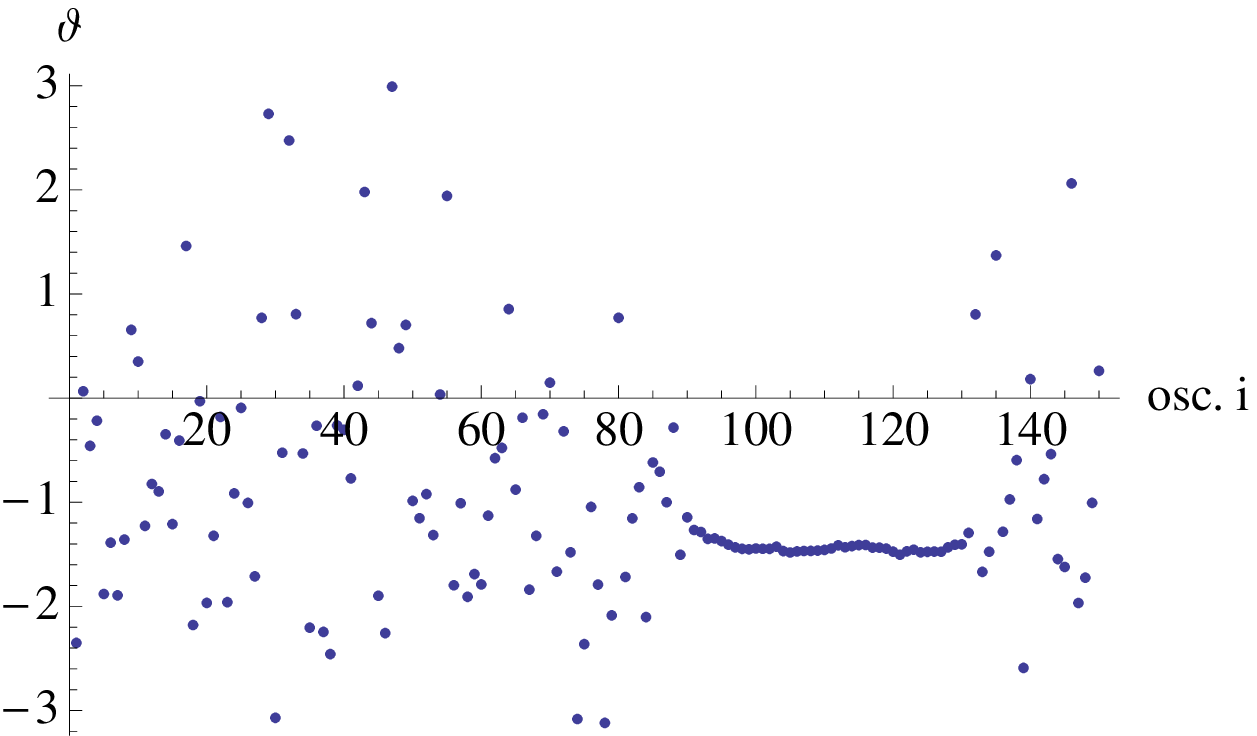}\\
\includegraphics[width=8cm]{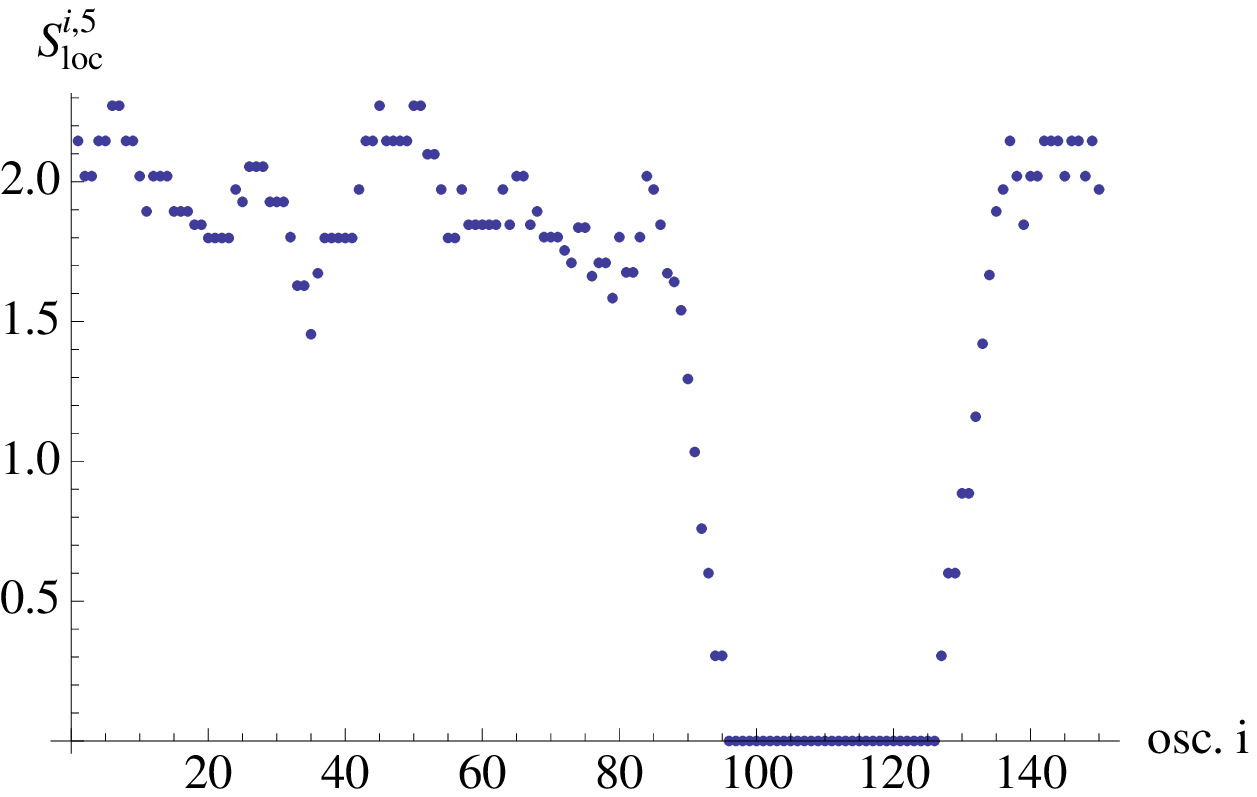}
\caption{\label{clchim} Phase snapshot of a classical chimera state of the oscillator chain defined by eq. \ref{clchimere} (up) and local classical entropy of the oscillator chain (down). The parameters are $N=150$ (number of oscillators), $\varpi=0$, $M=45$ (range of coupling), and $\alpha=1.46$. The initial condition consists to random values of $\vartheta_i(0)$ uniformly chosen in $[0,2\pi]$, the snapshot is taken at time $t=100 \frac{1}{\nu}$. The local classical entropy is defined by $S^{i,R}_{loc} = - \sum_{n=1}^{n_{res}} p_n^{i,R} \ln p_n^{i,R}$ where $p_n^{i,R}$ is the fraction of the $2R+1$ oscillators around the $i$-th one which are in the $n$-th microstate of a partition of $[-\pi,\pi]$ into $n_{res}$ microstates (the $n$-th microstate is the interval $[-\pi+(n-1)\frac{2\pi}{n_{res}},-\pi+n\frac{2\pi}{n_{res}}]$). We have chosen a resolution parameter $n_{res} = 20$ and a neighbourhood radius $R=5$.  $S^{i,R}_{loc}$ is the Shanon entropy (the disorder measure) of the chain piece of $2R+1$ oscillators centered on the $i$-th one.} 
\end{center}
\end{figure}

Recently intriguing states exhibiting both ordered and disordered dynamics have been discovered in long range coupled sets of oscillators \cite{cl_chimeria1} and have been highlighted in other coupled sets of classical (mechanical, electronic or opto-electronic) systems \cite{cl_chimeria2, cl_chimeria3, cl_chimeria4}. The particularity of such states is a part of the oscillator set exhibits an ordered dynamics (synchronized oscillations) whereas another part exhibits a disordered dynamics (desynchronized oscillations without correlation between the oscillators). This regime is not transient during a short time, these intriguing states have got a long and sometimes an infinite life duration: disorder does not spread to the whole set and the disordered part does not collapse to synchronized oscillations in a short time (see \cite{chimeria_chaine_finie} to find a discussion concerning the life duration of chimera states). These states have been called chimera, in reference to the mythological creature hybrid of a lion, a snake and a goat. Some studies have shown \cite{chimeria_chaine_finie, cl_chimeria5, cl_chimeria6} that for finite size chain, the disordered part presents a chaotic behaviour. An interesting simple example of chimera states have been studied in \cite{chimeria_chaine_finie}. It consists of a closed chain of $N$ oscillators with long range coupling of their phases:
\begin{equation}
\label{clchimere}
\dot \vartheta_i(t) = \varpi - \frac{\nu}{2M} \sum_{j=i-M}^{j+M} \sin(\vartheta_i(t)-\vartheta_j(t)+\alpha)
\end{equation}
where $\vartheta_i$ is the phase of the $i$-th oscillator, $\varpi$ and $\nu$ are constant frequencies, $\alpha$ is a constant angle and $M \in \{2,...,N/2-1\}$ is the range of coupling (the indices are taken modulo $N$). An example of a chimera state of this system is given fig. \ref{clchim}.
\\
In this paper we show that a simple quantum system, a closed chain of spins, offers quantum analogues of the chimera states if the couplings between the spins have the same structure that the couplings of the classical models. It is well known that spin chains can exhibit kinds of quantum disorder and of quantum chaos \cite{spinchain1, spinchain2, spinchain3, spinchain4}, and that quantum synchronization is related to the entanglement \cite{Orth, Manzano, Lee, Walter}. To involve a kind of chimera states, our model consists of a non-hermitian spin chain \cite{Bilstein, Castro, Bytsko, Schwager} which can be assimilated to a spin chain in contact with an environment. This model is presented next section. Disorder and chaos in the model are discussed in the following sections. These notions, which are ambiguous in quantum mechanics, can be enlightened by our model.

\section{The model}
We consider a closed chain of $N$ spins $\frac{1}{2}$. Let $\{\hat I_i\}_{i=1,...,N}$ be the set of the observables defined by
\begin{eqnarray}
\hat I_i & = & \frac{\hbar \omega_i}{2} \sigma_{zi} + \frac{\hbar \nu}{2 M} \sin \alpha \nonumber \\
& & + \frac{\hbar \nu}{2 M} \cos \alpha \sum_{j=i-M,j\not=i}^{i+M} (\sigma_{+i} \otimes \sigma_{zj} - \sigma_{zi} \otimes \sigma_{+j}) \nonumber \\
& & + \frac{\hbar \nu}{2 M} \sin \alpha \sum_{j=i-M,j\not=i}^{i+M} (\sigma_{+i} \otimes \sigma_{+j} + \sigma_{zi} \otimes \sigma_{zj}) \nonumber \\
\end{eqnarray}
where $\{\sigma_x,\sigma_y,\sigma_z\}$ are the Pauli matrices and $\sigma_\pm = \sigma_x \pm \imath \sigma_y$ (the indices denote the spin on which the Pauli matrix acts as an operator; the indices are taken modulo $N$). $M \in \{2,...,N/2-1\}$ is the range of coupling between the spins, $\nu$ is a constant frequency, $\alpha$ is a constant angle and $\omega_i$ is the Larmor frequency of the $i$-th spin in a local magnetic field. The observable $\hat I_i$ is a quantum analogue of the equation (\ref{clchimere}) of the classical model. Indeed let \mbox{$|\theta,\phi\rangle = \cos \frac{\theta}{2}|\uparrow \rangle + e^{\imath \phi} \sin \frac{\theta}{2} |\downarrow \rangle$} be the spin coherent state \cite{Perelomov}, i.e. the quantum state closer to the classical spin state defined by the phase space point $(\theta,\phi)$ ($\theta$ and $\phi$ are the angles, which are the coordinates on the Bloch sphere). We have $\langle \theta,\phi|\sigma_z|\theta,\phi \rangle = \cos \theta$ and $\langle \theta,\phi|\sigma_+|\theta,\phi\rangle = e^{\imath \phi} \sin \theta$. Let $|\underline \theta \rangle = |\theta_1,0\rangle \otimes .... \otimes |\theta_N,0\rangle$ be the coherent state for the $N$ spins of the chain with $\phi_1=...=\phi_N=0$. We have
\begin{equation}
\label{intpremqu}
\langle \underline \theta| \hat I_i |\underline \theta \rangle = \frac{\hbar \omega_i \cos \theta_i}{2} + \frac{\hbar \nu}{2M} \sum_{j=i-M}^{j+M} \sin(\theta_i-\theta_j+\alpha)
\end{equation}
which is similar to the classical first integral $I_i = \dot \vartheta_i + \frac{\nu}{2M} \sum_{j=i-M}^{j+M} \sin(\vartheta_i-\vartheta_j+\alpha)$ with respect to the structure of the couplings (we can note that the free terms,  $\frac{\omega_i \cos \theta_i}{2}$ and $\dot \vartheta_i$ are completly different because the individual elements of each chain have a different physical nature, quantum spins for $\hat I_i$ and classical oscillators for $I_i$). We note that this is only a mathematical analogy, the quantum model is not the quantization of the classical one, and the two models concern different physical systems (a set of oscillators for the classical one, and a set of spins for the quantum one). We note that $[\hat I_i,\hat I_j] \not= \delta_{ij}$ like $\{I_i,I_j\} \not= \delta_{ij}$ (with $\{.,.\}$ the Poisson bracket). We use the observables $\{\hat I_i\}_i$ to define the Hamiltonian of our quantum analogue of the system (\ref{clchimere}):
\begin{eqnarray}
\label{ham}
H & = & \sum_{i=1}^N \hat I_i \\
& = & \sum_{i=1}^N \frac{\hbar \omega_i}{2} \sigma_{zi} + \frac{N \hbar \nu}{2 M} \sin \alpha \nonumber \\
& &  + \frac{\hbar \nu \sin \alpha}{2 M} \sum_{i=1}^N\sum_{\begin{array}{c} \vspace{-0.5em} \scriptstyle j=i-M \\ \scriptstyle j\not=i \end{array}}^{i+M} (\sigma_{+i} \otimes \sigma_{+j} + \sigma_{zi} \otimes \sigma_{zj}) \nonumber \\
\end{eqnarray}
Note that this quantum system is analogue to the system (\ref{clchimere}) in the sense of (\ref{intpremqu}), i.e. the couplings have the same structure. However it is not the quantization of the model (\ref{clchimere}) and this last one is not the classical limit of the system (\ref{ham}). A discussion about the semi-classical model of the system (\ref{clchimere}) can be found in \cite{Hermoso}, but in this paper we want to exhibit purely quantum chimera states with a spin chain model. $H$ is not hermitian but its spectrum is real since the matrix representation of $H$ is upper diagonal with real values on the diagonal ($\pm \frac{\hbar \omega_i}{2} + k \frac{\hbar \nu}{2M} \sin \alpha$, $k\in \mathbb Z$). We say that $H$ is quasi-hermitian\footnote{Note that $H$ is quasi-hermitian in the sense of \cite{Bytsko} only if $\omega_i=0$ $\forall i$: $H^\dagger= \eta H \eta^{-1}$ with $\eta = \sigma_{x1} \otimes ... \otimes \sigma_{xN}$ ($\sigma_x \sigma_- \sigma_x = \sigma_+$ and $\sigma_x \sigma_z \sigma_x = - \sigma_z$).}. $H$ can be viewed as the effective Hamiltonian of a long range coupling Heisenberg spin chain in contact with an environment (examples of non-hermitian effective Hamiltonians can be found in \cite{Lee1, Lee2}).

\section{Quantum chimera states}
\begin{figure}
\begin{center}
\includegraphics[width=7cm]{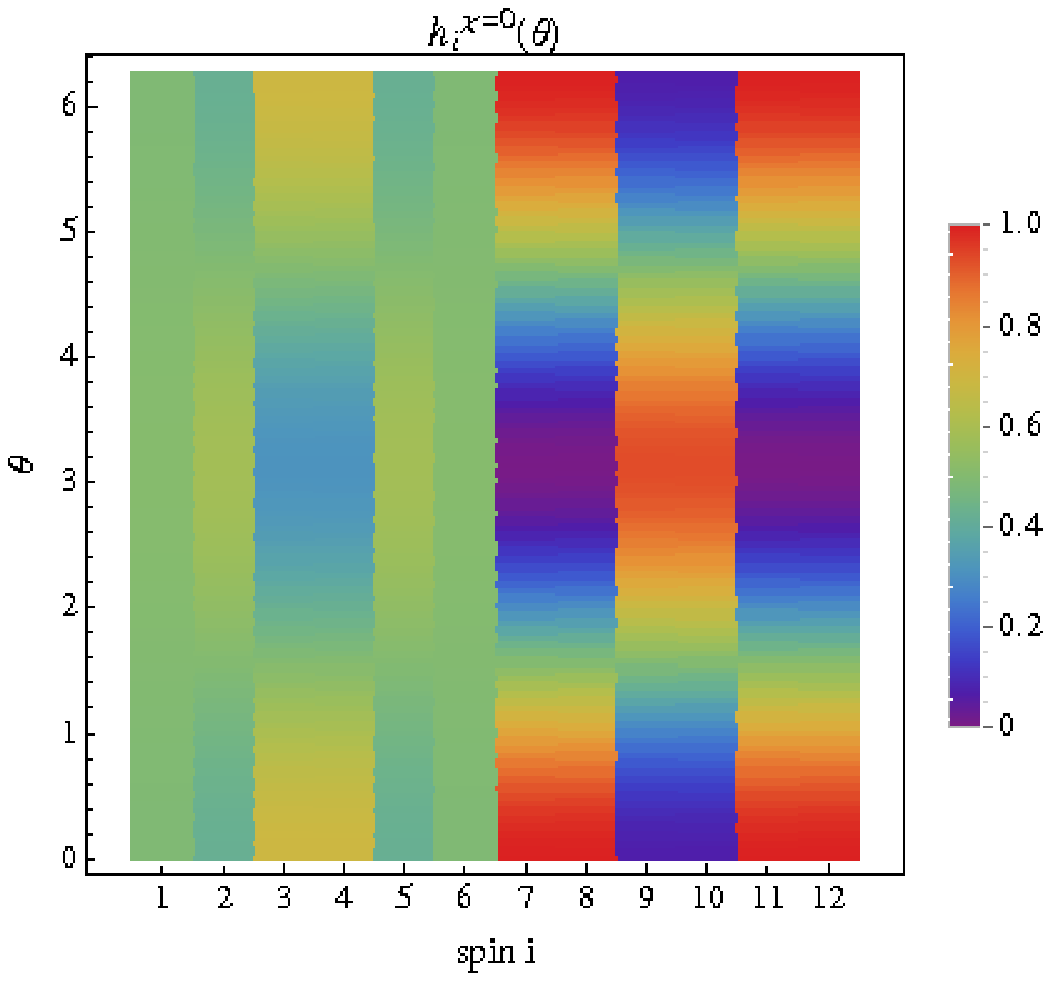}\\
\includegraphics[width=8.5cm]{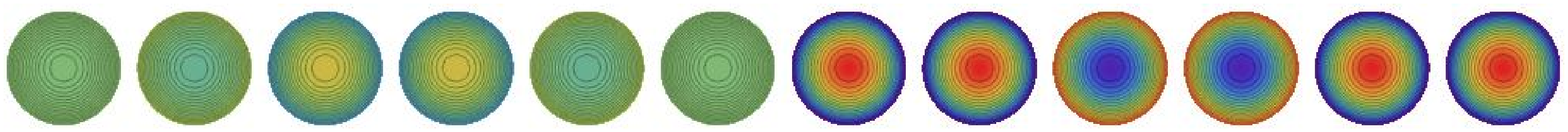}\\
\includegraphics[width=8cm]{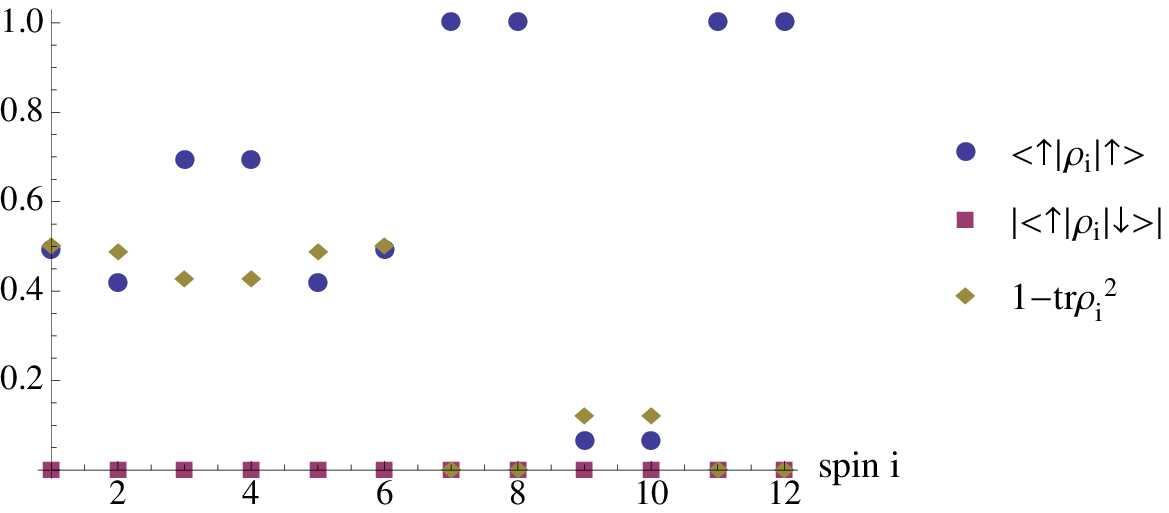}\\
\caption{\label{chimeria1} Husimi distribution $h_i^{\chi_n}(\theta,0) = |\langle \theta,0|\rho_i^{\chi_n}|\theta,0\rangle|$ (top figure); $h_i^{\chi_n}(\theta,\phi) = |\langle \theta,\phi|\rho_i^{\chi_n}|\theta,\phi\rangle|$ (middle figure) where the Bloch sphere of each spin is represented by azimuthal projection centered on the north pole ($\theta=0$) the limit circle being the south pole ($\theta=\frac{\pi}{2}$); and up populations, coherences and linear entropies (bottom figure) of the spins of the chain in the eigenstate $|\chi=0\rangle$ with $N=12$, $M=3$, $\omega_1=...=\omega_N=1.\ a.u.$, $\nu=1\ a.u.$, $\alpha = \frac{\pi}{2}$ ($a.u.$: atomic units).}
\end{center}
\end{figure}

\begin{figure}
\begin{center}
\includegraphics[width=7cm]{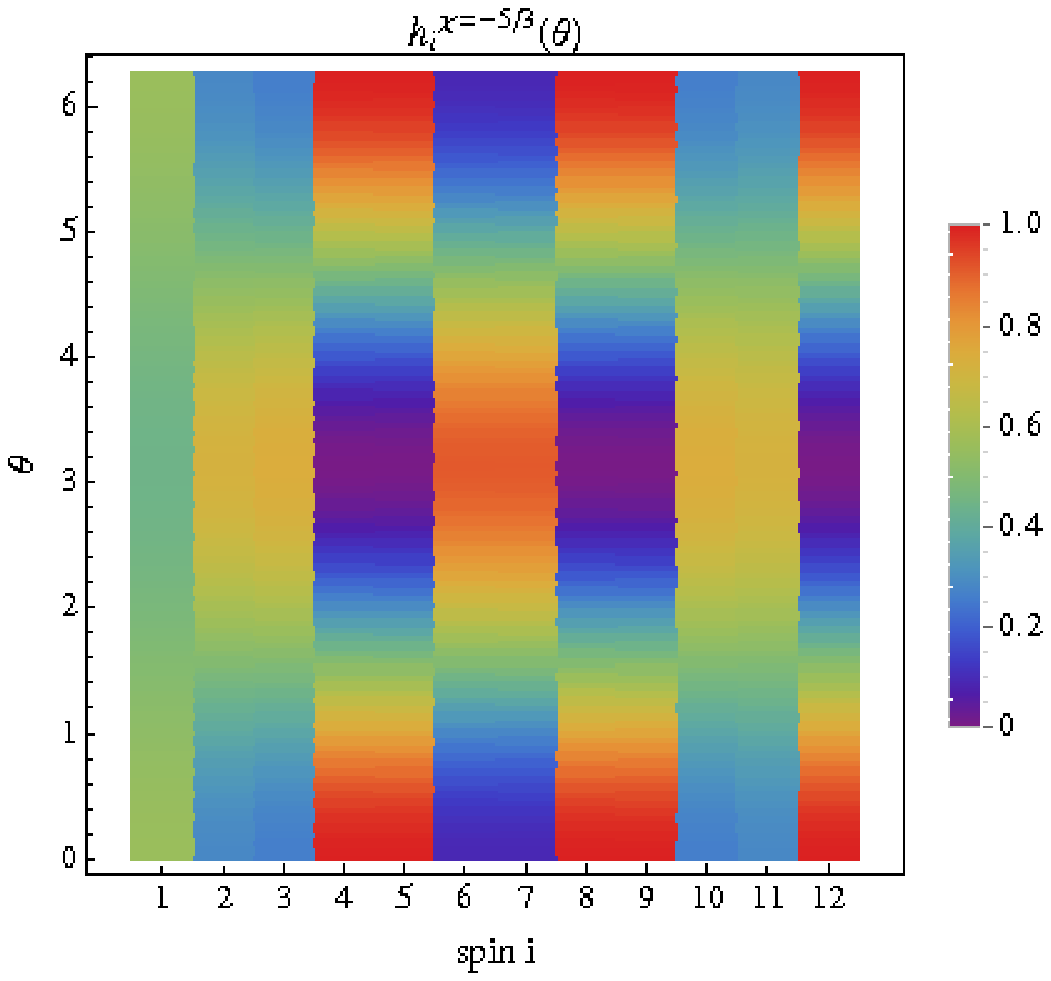}\\
\includegraphics[width=8.5cm]{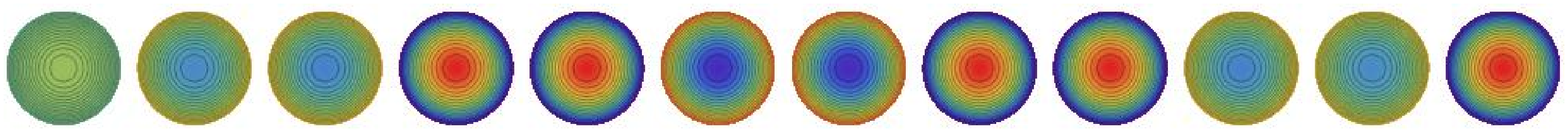}\\
\includegraphics[width=8cm]{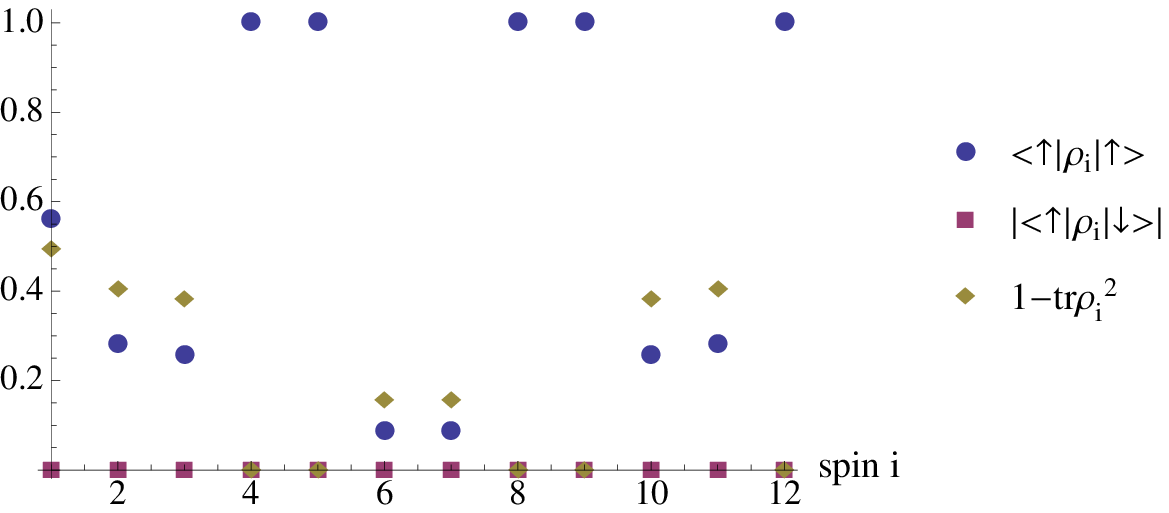}\\
\caption{\label{chimeria2} Same as figure \ref{chimeria1} with the eigenstate $|\chi=-5/3\rangle$.}
\end{center}
\end{figure}

We consider the eigenstates  and the biorthogonal eigenstates of $H$ respectively:
\begin{equation}
H|\chi_n \rangle = \chi_n |\chi_n \rangle \qquad H^\dagger|\chi_n^\sharp \rangle = \chi_n |\chi_n^\sharp \rangle
\end{equation}
with $\chi_n \in \mathbb R$ and $\langle \chi_n^\sharp|\chi_p\rangle = \delta_{np}$. In order to enlighten the similarity of these eigenstates with chimera states, we consider the Husimi distribution \cite{Husimi} $h_i^{\chi_n}(\theta,\phi) = |\langle \theta,\phi|\rho_i^{\chi_n}|\theta,\phi\rangle|$ where $\rho_i^{\chi_n} = \tr_i |\chi_n\rangle \langle \chi_n|$ is the density matrix of the spin $i$ when the chain is in the state $|\chi_n\rangle$ ($\tr_i$ denotes the partial trace over all spin spaces except the $i$-th one). $h_i^{\chi_n}(\theta)$ measures the probability of similarity between the mixed quantum state $\rho_i^{\chi_n}$ and the classical spin state characterized by an angle $\theta$ with the $z$-axis. To complete the analysis we consider also the up population $p_i^{\chi_n} = \langle \uparrow|\rho_i^{\chi_n}|\uparrow\rangle$ (the occupation probability of the state up by the spin $i$), the coherence of the spin $i$ $c_i^{\chi_n} = |\langle \uparrow|\rho_i^{\chi_n}|\downarrow\rangle|$, and the linear entropy $S_i^{\chi_n} = 1 - \tr (\rho_i^{\chi_n})^2$ (the entanglement measure of the spin $i$ with the other spins).\\
A typical eigenstate is shown figure \ref{chimeria1}.
We observe its similarity with the classical states of the model (\ref{clchimere}) studied in \cite{chimeria_chaine_finie} and fig. \ref{clchim}: a part of the spin chain (from the spin 1 to the spin 6) presents a large entropy and the other one, a zero (or a small) entropy. The similarity with fig. \ref{clchim} is obvious, in the presented classical chimera state, a part of the oscillator chain (from the oscillator 95 to the oscillator 125) presents a small local entropy, whereas the rest of the chain presents a large entropy. But in contrast with the classical case where the entropy measures the disorder, in this quantum context the entropy measures the entanglement. In comparison, the computation of the same quantities for different models of chaotic or random spin chains or glasses \cite{spinchain1, spinchain2, spinchain3, spinchain4, modelmixte} shows eigenstates with a large entanglement which is uniform on the chain (or with small variations between nearest neighbour spins). These models do not involve states with both some spins highly entangled and the other ones totally not entangled as in figure \ref{chimeria1}. The ``vertical green region'' of the Husimi distribution (the entangled region from spin 1 to spin 6, which is also characterised by a large entropy and a zero coherence in the down part of figure \ref{chimeria1}) corresponds to the ``decoherent part'' of the chain and the region where the Husimi distribution shows spins ``aligned'' with the up or the down directions (the non entangled region from spin 7 to spin 12, which is also characterised by a small entropy and a population close to 0 or 1 in the down part of figure \ref{chimeria1}) corresponds to the coherent part of the chain. The chain is closed and other eigenstates present an entangled region centered on other spins. Moreover, in contrast with the classical case, the green region is not necessarily connected as in figure \ref{chimeria2}.
The quantum states like figures \ref{chimeria1} and \ref{chimeria2} can be considered as quantum chimera states. Note that the present model like the chaotic or the random models \cite{spinchain1, spinchain2, spinchain3, spinchain4, modelmixte} presents also totally regular (non entangled) states (as classical chimera states coexist with fully synchronized states).\\
Remark: we can note that the considered chimera states figures \ref{chimeria1} and \ref{chimeria2} are stationary quantum states (eigenvectors of $H$). Classical chimera states have a dynamical character, however they have a long life duration. They correspond then to a metastable regime (until the collapses studied in \cite{chimeria_chaine_finie} for finite size chains) and not to a short life transient regime. Moreover for some classical models of infinite chains \cite{cl_chimeria1}, the classical chimera states present an infinite life duration and correspond to a permanent regime. In quantum mechanics, classical permanent regimes become quantum stationary eigenstates. For example, the classical permanent regimes of the harmonic oscillator (sinusoidal oscillations with respect to the time) become stationary quantum eigenstates (product of an Hermite polynomial with an exponential with respect to the displacement); and the classical permanent regimes of the classical Keplerian problem (elliptic orbits followed with respect to the time) becomes the stationary hydrogen wavefunctions. The dynamical character of the permanent classical regimes becomes the non-local character of the stationary quantum eigenstates (spatial delocalisation, entanglement, etc). In contrast, transient regime states are time-dependent both for classical and quantum dynamics. The classical metastable regimes (regimes with a long life duration but not infinite) seem to become quantum stationary eigenstates but associated with a non-hermiticity (the left eigenvectors are different from the right ones, in accordance with the difference concerning the evolutions to the future and to the past induced by the irreversible collapse or destabilisation of the state at the end of its life duration). For example to make appear quantum resonances in the spectral properties, it is necessary to use a complex dilation or an optical potential which induces that the Hamiltonian becomes non-hermitian (see \cite{Moiseyev}). It is then not surprising for a finite chain of spins, to find analogues of long life duration classical chimera states as quantum stationary eigenstates associated with the non-hermiticity of the Hamiltonian. The dynamical character of the classical chimera states is replaced by the non-local character of the entanglement of the spins (green regions in figures \ref{chimeria1} and \ref{chimeria2}).

\section{Disorder and entanglement}
\begin{figure}
\begin{center}
\includegraphics[width=6.cm]{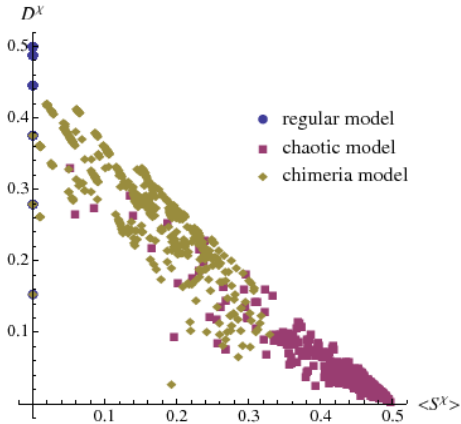}\\
\caption{\label{disorder} Distribution in the plane (entanglement $\langle S^{\chi} \rangle$ - disorder $D^{\chi}$) of a representative sample of eigenstates of the chimera model (\ref{ham}) (with $N=12$, $M=3$, $\omega_i= 0.2+i\frac{0.4}{N}\ u.a.$, $\nu=1\ a.u.$ and $\alpha=1.46$ -- $a.u.$: atomic units --), of eigenstates of a totally regular model (an Ising-Z spin chain with nearest neighbour interactions), and of the totally chaotic model studied in \cite{spinchain3, spinchain4}.}
\end{center}
\end{figure}

Disorder does not have the same status for quantum or classical systems. It is the entanglement which is involved by the quantum decoherence and not the disorder. It must be interesting to measure these two physical concepts globally. The average linear entropy $\langle S^{\chi_n} \rangle = \frac{1}{N} \sum_{i=1}^N S^{\chi_n}_i$ is a measure of the mean entanglement of the chain in the state $\chi_n$. If each spin is in a pure state, the linear entropy $1 - \tr\langle \rho^{\chi_n} \rangle^2$  of the average state $\langle \rho^{\chi_n} \rangle = \frac{1}{N} \sum_{i=1}^N \rho^{\chi_n}_i$ is a measure of the disorder because it is zero if all the pure states are equal and is large if the pure states are strongly different. But if the spins are in mixed states, $1 - \tr\langle \rho^{\chi_n} \rangle^2$ includes also the entanglement entropy of the chain. We propose then  as a measure of the quantum disorder $D^{\chi_n} = 1 - \tr\langle \rho^{\chi_n} \rangle^2 - \langle S^{\chi_n} \rangle$. We have represented figure \ref{disorder} the entanglement and  disorder distribution for the chimera model in comparison with chaotic and regular models.
The totally regular systems present eigenstates concentrated on the zero entanglement axis (the largest disordered states being with half of the spins in the pure state up and the other half in the pure state down). The chaotic systems present eigenstates essentially concentrated in the region with small disorder and large entanglement. The chimera model presents a distribution of its eigenvectors clearly between these two cases, characterizing its hybrid nature.

\section{Chaotic behaviour}
\begin{figure}
\begin{center}
\includegraphics[width=6.5cm]{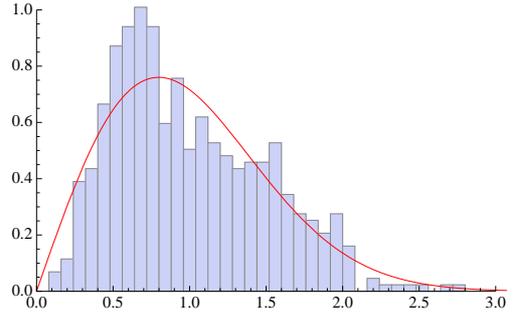}\\
\caption{\label{LSD} Level spacing distribution (LSD) of the spectrum (after removing the degeneracies) of the chimera Hamiltonian (\ref{ham}) (with $N=12$, $M=3$, $\omega_i = 0.2+i\frac{0.4}{N}\ u.a.$, $\nu=1\ a.u.$, $\alpha=1.46$ -- $a.u.$ : atomic units --) and the profile of a Wigner-Dyson distribution (red continuous curve).}
\end{center}
\end{figure}

\begin{figure}
\begin{center}
\includegraphics[width=6.5cm]{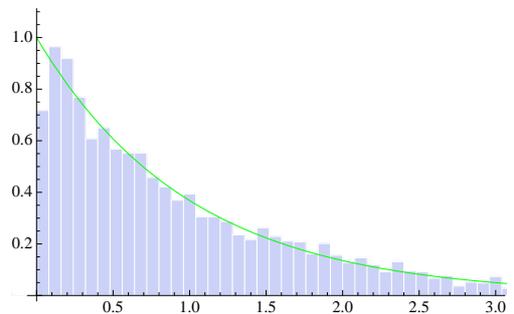}\\
\caption{\label{LSD2} Level spacing distribution (LSD) of the spectrum of the chimera Hamiltonian (\ref{ham}) (with $N=12$, $M=3$, $\omega_i = 0.2+0.4 \sin(\frac{i^2}{N^2} \frac{\pi}{2})\ u.a.$, $\nu=1\ a.u.$, $\alpha=1.46$ -- $a.u.$ : atomic units --) and the profile of a Poisson distribution (green continuous curve).}
\end{center}
\end{figure}

\begin{figure}
\begin{center}
\includegraphics[width=6.5cm]{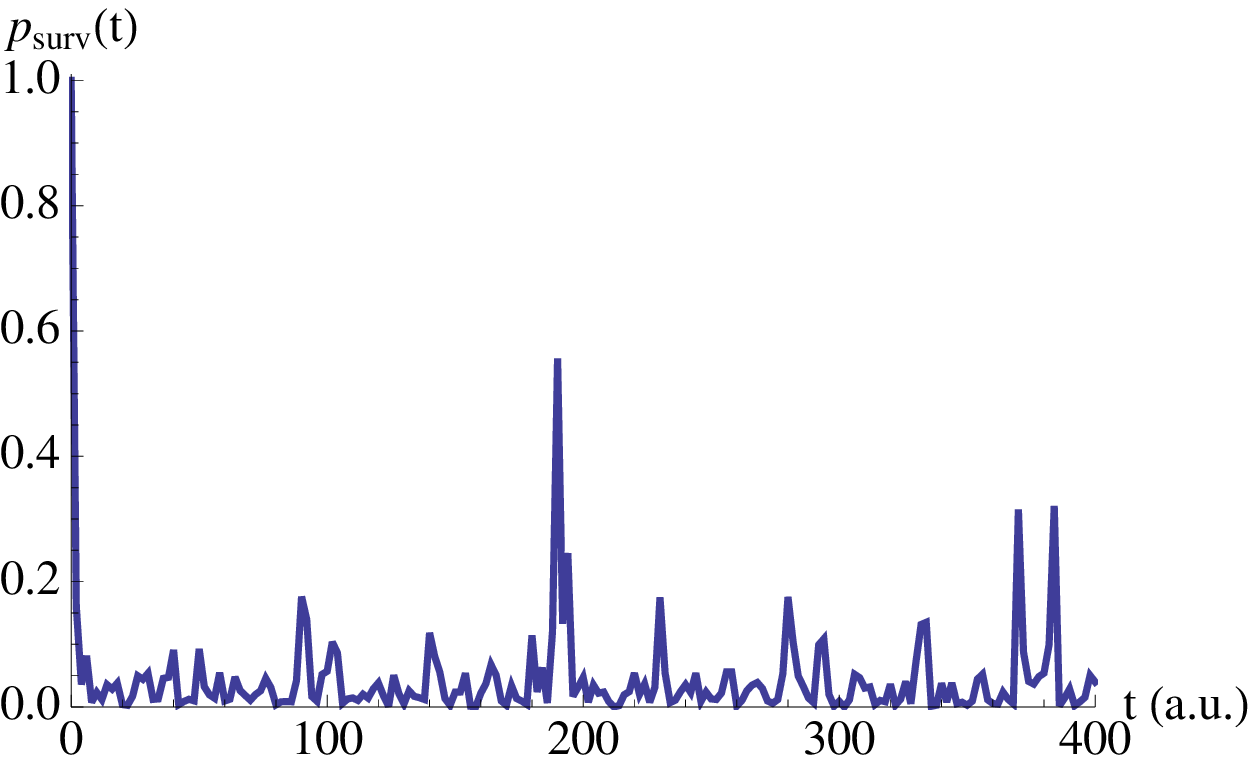}\\
\includegraphics[width=6.5cm]{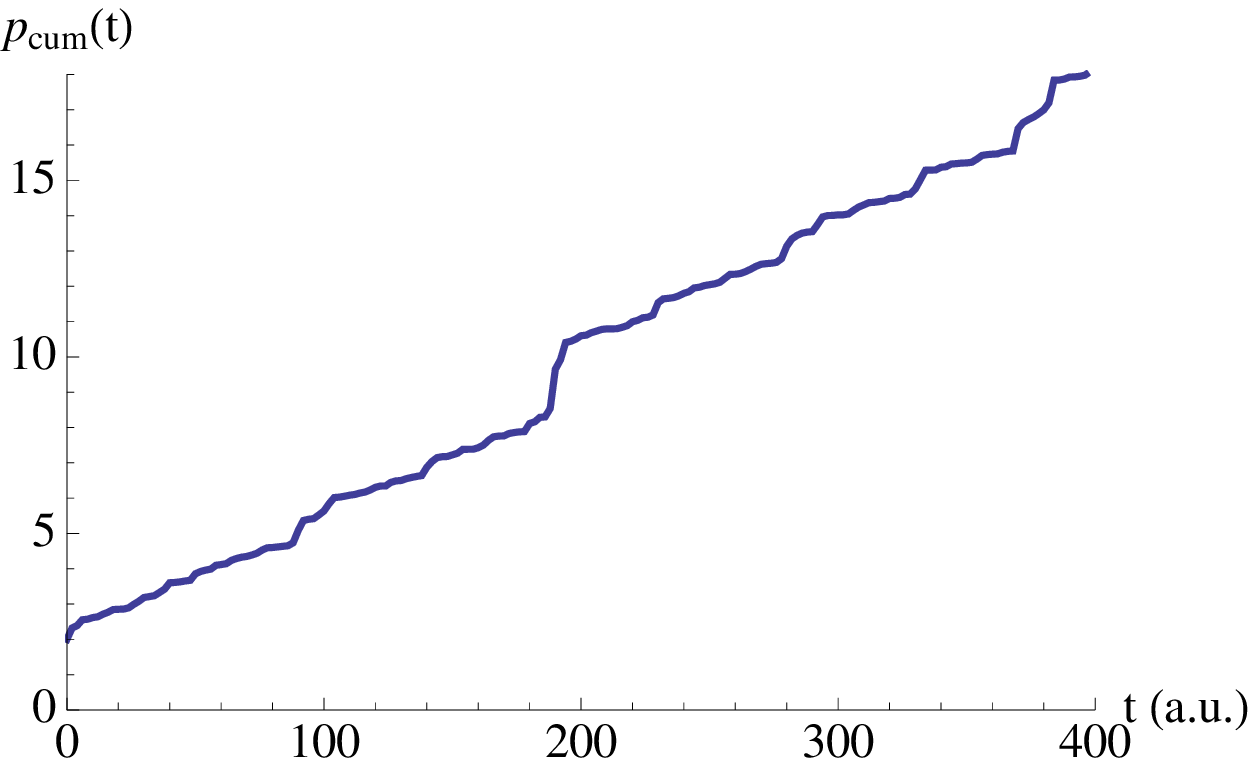}\\
\caption{\label{proba} Survival probability and cumulated survival probability of the state  $|\downarrow \uparrow \downarrow \downarrow \downarrow \uparrow \downarrow \downarrow \downarrow \downarrow \uparrow \downarrow \rangle$ with respect to the time (with $N=12$, $M=3$, $\omega_i = 0.2+i\frac{0.4}{N}\ u.a.$, $\nu=1\ a.u.$ and $\alpha=1.46$ -- $a.u.$ : atomic units--).}
\end{center}
\end{figure}

\begin{figure}
\begin{center}
\includegraphics[width=6.5cm]{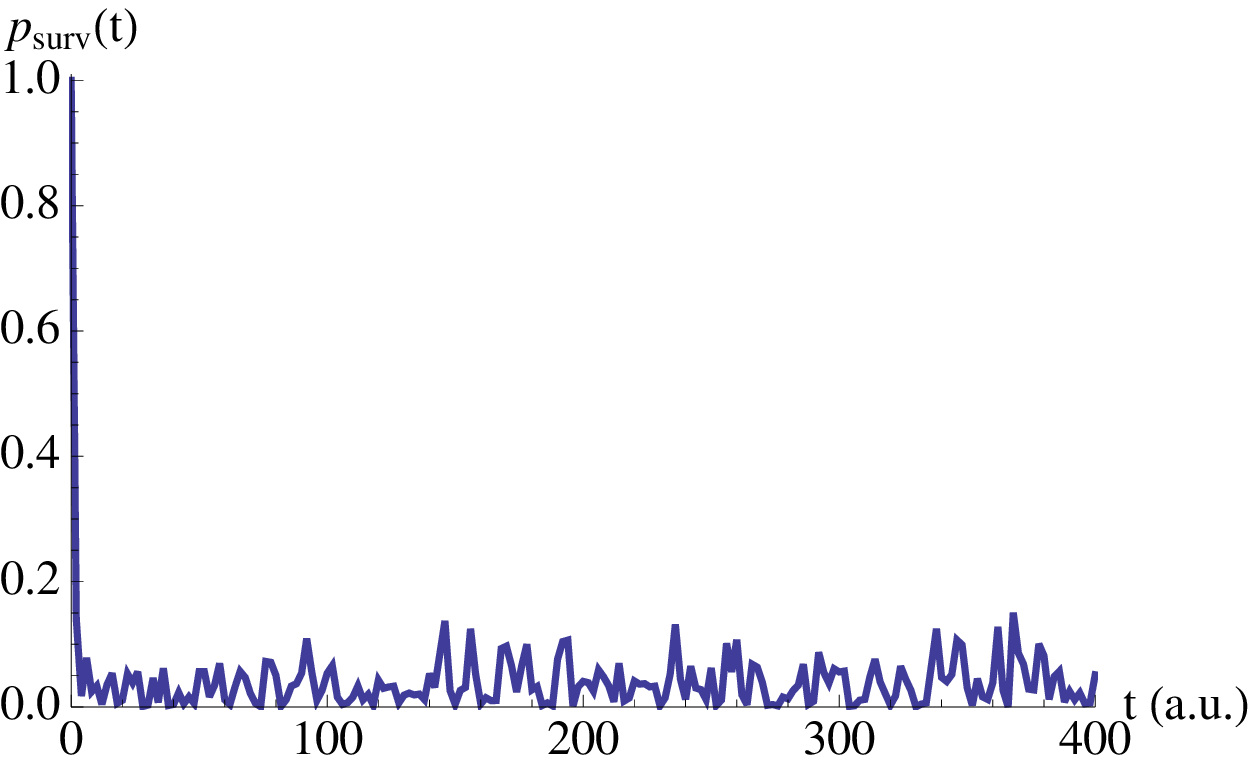}\\
\includegraphics[width=6.5cm]{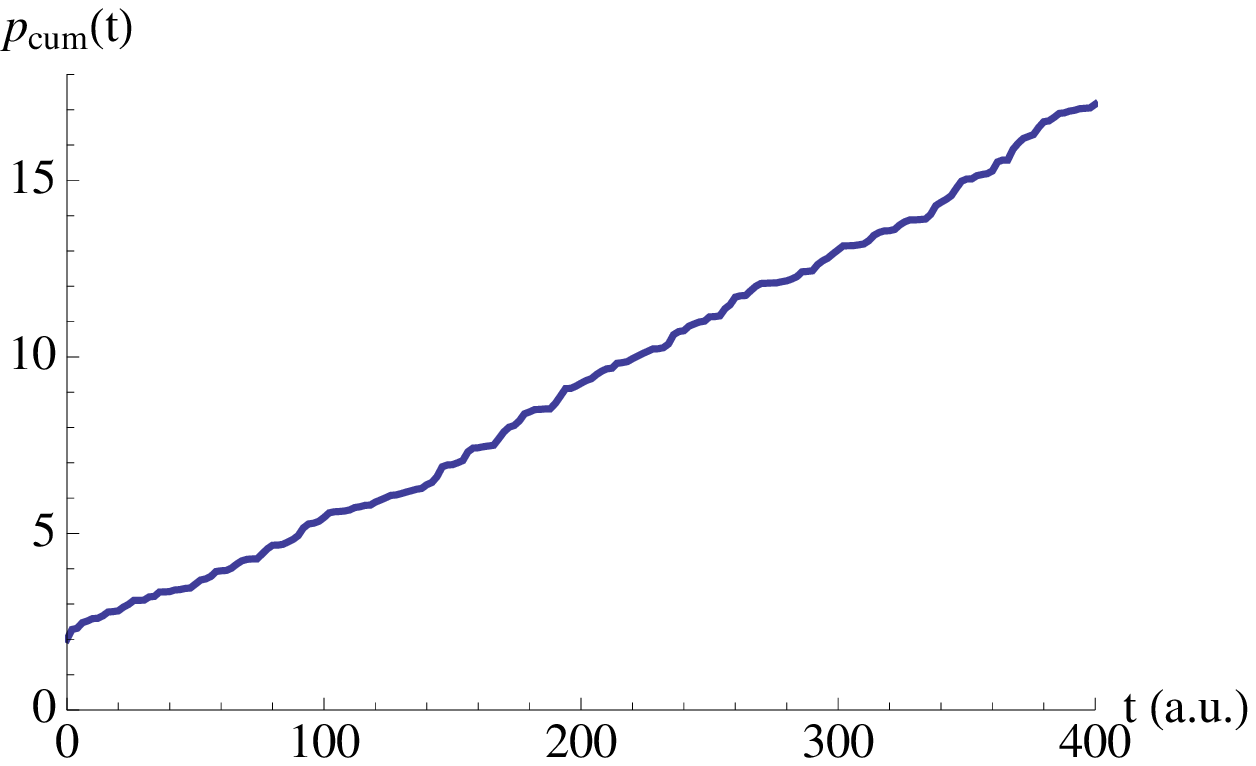}\\
\caption{\label{proba2} Survival probability and cumulated survival probability of the state  $|\downarrow \uparrow \downarrow \downarrow \downarrow \uparrow \downarrow \downarrow \downarrow \downarrow \uparrow \downarrow \rangle$ with respect to the time (with $N=12$, $M=3$, $\omega_i = 0.2+0.4 \sin(\frac{i^2}{N^2} \frac{\pi}{2})\ u.a.$, $\nu=1\ a.u.$ and $\alpha=1.46$ -- $a.u.$ : atomic units--).}
\end{center}
\end{figure}

\begin{figure}
\begin{center}
\includegraphics[width=6.5cm]{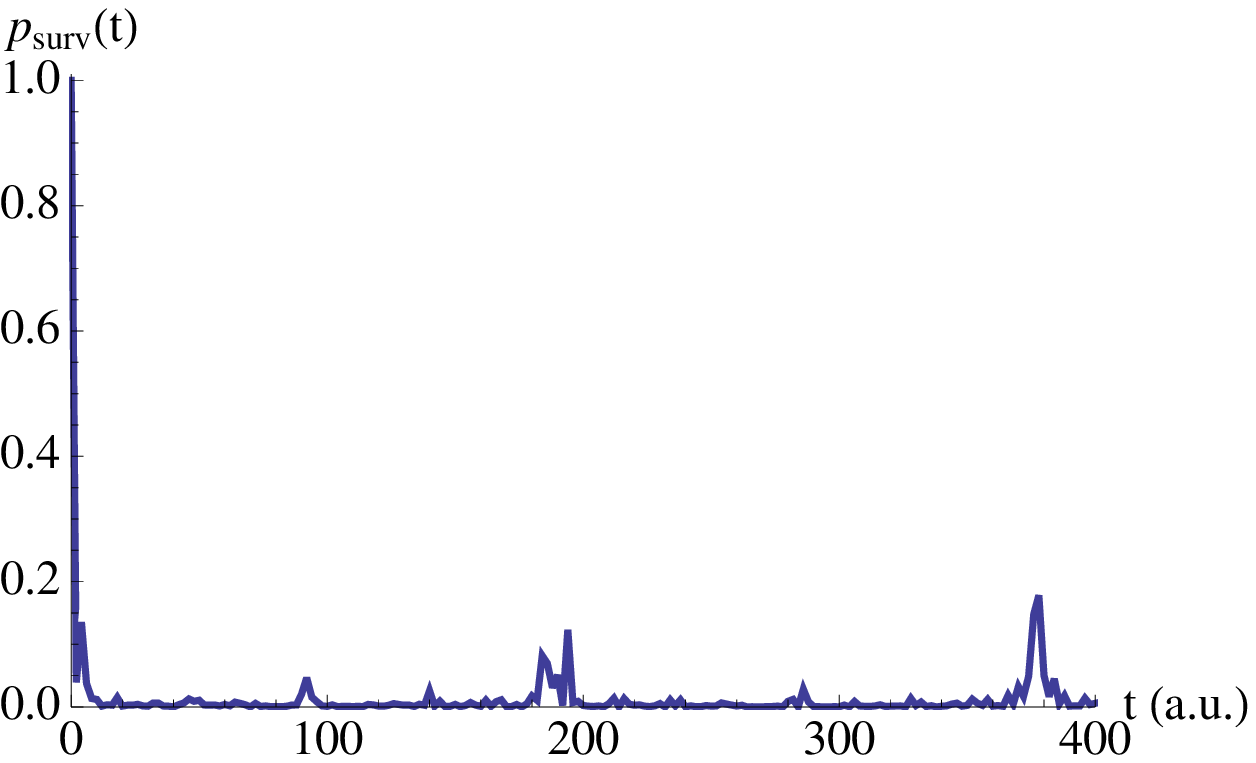}\\
\includegraphics[width=6.5cm]{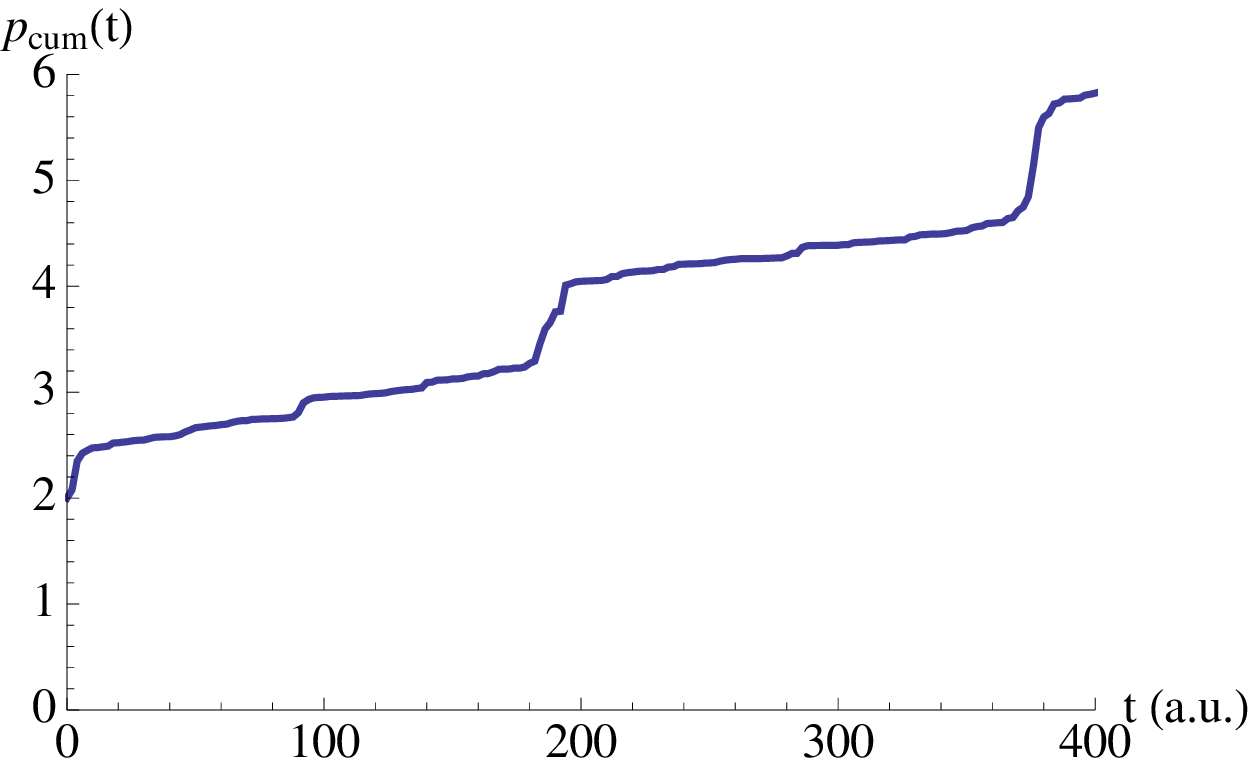}\\
\caption{\label{probchimera} Survival probabilities (up) and cumulated survival probability (down) of a chimera states of the form $|\downarrow\downarrow \downarrow \downarrow \downarrow \downarrow **** \downarrow \downarrow \rangle$ with $N=12$, $M=3$, $\omega_i = 0.2+i\frac{0.4}{N}\ u.a.$ , $\nu=1\ a.u.$ and $\alpha=1.46$ ($a.u.$ : atomic units).}
\end{center}
\end{figure}

\begin{figure}
\begin{center}
\includegraphics[width=6.5cm]{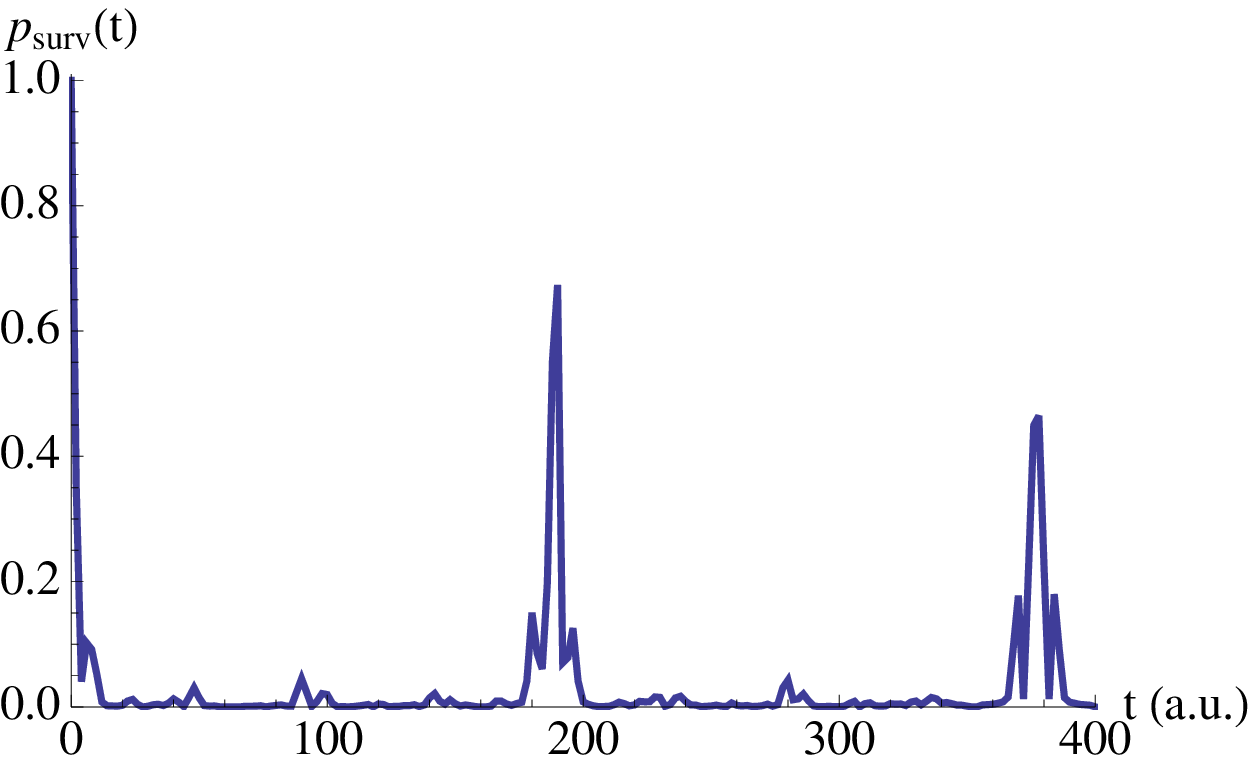}\\
\includegraphics[width=6.5cm]{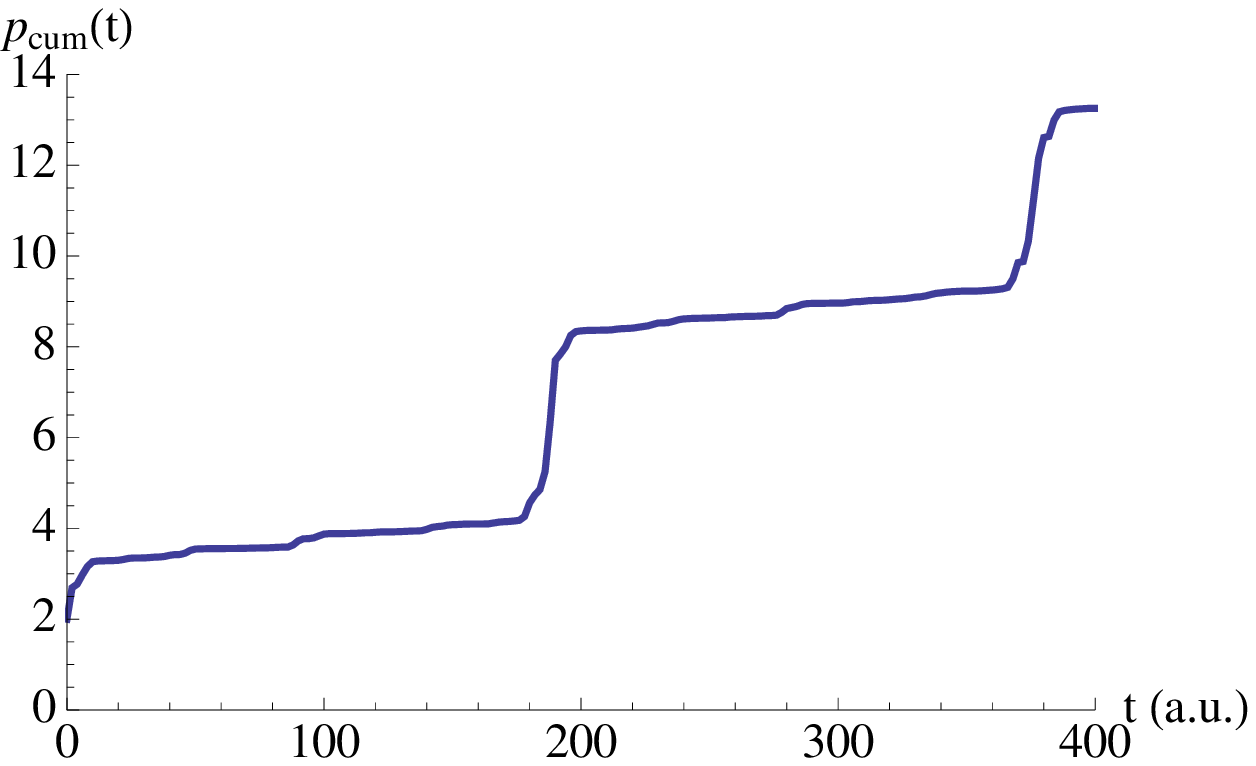}\\
\caption{\label{probchimera2} Survival probabilities (up) and cumulated survival probability (down) of a chimera states of the form $|\downarrow \downarrow * \uparrow \downarrow * \downarrow \uparrow \downarrow \downarrow \downarrow \uparrow \rangle$ with $N=12$, $M=3$, $\omega_i = 0.2+i\frac{0.4}{N}\ u.a.$ , $\nu=1\ a.u.$ and $\alpha=1.46$ ($a.u.$ : atomic units).}
\end{center}
\end{figure}

A last question concerns the chaotic nature of the quantum chimera model. Quantum chaos is an ambiguous concept since in classical dynamics the chaos is strongly linked to the non-linear effects whereas the quantum dynamics is fundamentally a linear theory. A commonly used criterion of quantum chaos for spin systems is the level spacing distribution (LSD) of the spectrum \cite{spinchain1, spinchain2, spinchain3, spinchain4}. A regular system presents a LSD as Dirac peaks, a (pseudo)-random system presents a LSD as a Poisson distribution (characterizing the disorder of the energy levels without correlation) and a chaotic system presents a LSD as a Wigner-Dyson distribution (characterizing the disorder of the energy levels with correlations). The chimera system (\ref{ham}) presents a LSD which seems to be Dirac peaks if $\omega_1=...=\omega_N$. But with $\omega_i=0.2+i\frac{0.4}{N}\ a.u.$ the LSD\footnote{The LSD is computed by using the following unfolding procedure. We consider $\{(\chi_n,n)\}_{n=1,...,2^N}$ where the eigenvalue labels are ordered such that $n>p \Rightarrow \chi_n > \chi_p$. We consider $\chi \mapsto f(\chi)$ the function obtained by spline interpolation of $\{(\chi_{jp+1},jp+1)\}_{j=0,...,2^N/p-1}$ where $p\in \mathbb N^*$ is an integer (in the figure we have $p=10$). The unfolding spectrum $\{f(\chi_n)\}_{n=1,...,2^N}$ has a mean level spacing around 1.} seems not trivial as shown figure \ref{LSD}.
After removing the many degeneracies in the spectrum, the LSD is roughly similar to a Wigner-Dyson distribution characteristic of quantum chaos.  In other cases, for example with $\omega_i=0.2+0.4 \sin(\frac{i^2}{N^2} \frac{\pi}{2})\ a.u.$ the LSD (figure \ref{LSD2}) seems to be a Poisson distribution (in this case, no degeneracy occurs).
The LSD is then highly dependent on the distribution of the Larmor frequencies $\{\omega_i\}$, the chaotic character of the quantum chimera states is possibly different with respect to the choice of this distribution.\\
Another criterion of quantum chaos \cite{quchaos} concerns the dynamical behavior of a chosen state $\psi_0$ with respect to its survival probability $p_{surv}(t) = |\langle \psi_0|e^{-\ihbar^{-1} Ht}|\psi_0 \rangle_\sharp|^2$ ($\langle.|.\rangle_\sharp$ denotes the modified inner product associated with the biorthogonality \cite{Bytsko}, the so-called c-product \cite{Moiseyev}). $\psi_0$ is a bound state if its survival probability is constant, or presents periodic or quasiperiodic oscillations. $\psi_0$ is a scattering state if its survival probability falls quickly and definitively to zero. $\psi_0$ is a chaotic state if its survival probability chaotically oscillates with globally a slow decrease to zero with erratic resurgences of non-zero probabilities. \\ These behaviours can be enlightened by considering the cumulated survival probability $p_{cum}(t) = \int_0^t p_{surv}(t')dt'$. For a bound state the cumulated survival probability grows linearly, for a scattering state it quickly increases until a maximal value and then remains constant, for a chaotic state it grows on and on but not linearly. A chaotic quantum system is then a system exhibiting some chaotic states. 
Let $|\psi_0\rangle$ be a state with the spins in states up or down (without superposition) relatively disordered, for example\footnote{Note that $|\downarrow ... \downarrow \rangle$ is an eigenstate like for the totally chaotic models \cite{spinchain1, spinchain2, spinchain3, spinchain4}. But for these cases $|\downarrow ... \downarrow \uparrow \rangle$ presents a survival probability with a chaotic behaviour. This is not the case for the chimera model. Due its hybrid nature, it needs at least two distant turned spins to generate a chaotic behaviour of the survival probability otherwise the state is too close to an eigenstate where the turned spins are in the regular region. This question of the choice of the initial state to exhibit a chaotic behaviour in the quantum dynamics, can be compared with the limited range of initial conditions involving a chimera phenomenon in the classical systems.} $|\psi_0\rangle = |\downarrow \uparrow \downarrow \downarrow \downarrow \uparrow \downarrow \downarrow \downarrow \downarrow \uparrow \downarrow \rangle$. Such a state is close to a chaotic state for the case with a LSD being a Wigner-Dyson distribution, as shown by its survival probability and its cumulated survival probability drawn figure \ref{proba}.
We see that the survival probability seems to ``chaotically'' oscillate with a global decrease and with resurgences. The cumulated survival probability grows on and on with an almost linear growth. This ambiguous behaviour is certainly the manifestation of the nature of the chimera system which is a hybrid of a both chaotic and regular system. For the case with a LSD being a Poisson distribution, the survival probability and the cumulative probability are shown figure \ref{proba2}.
In this case the resurgences disappear, and the behaviour of the state is close to a scattering state but with a noise of non zero probabilities in the neighbourhood of zero. This behaviour is in accordance with the Poisson distribution of the LSD which is not characteristic to a quantum chaotic behaviour but to a quantum pseudo-random behaviour. In comparison with the classical case, if the model (\ref{clchimere}) presents chimera states which are always transient chaotic as shown in \cite{chimeria_chaine_finie}, the main characteristic of the classical chimera states are the coexistence of ordered and disordered dynamics in the same chain. The disordered dynamics is not necessary chaotic. The quantum dynamics observed figure \ref{proba2} seems to be characteristic of the disorder (by the presence of the noise) but without an obvious chaotic behaviour (no resurgence of the survival probability).\\
An interesting question is the dynamical behaviour of a chimera state. The survival probability of a chimera state $|\chi_n\rangle$ is trivial since it is a right eigenvector of $H$, but $H^\dagger$ presents also chimera states $|\chi_n^\sharp \rangle$ which do not generally coincide with those of $H$ (chimera left eigenvectors). Figures \ref{probchimera} and \ref{probchimera2} shows two examples of survival probability with chimera states of the form $|\psi_0\rangle = |\downarrow\downarrow \downarrow \downarrow \downarrow \downarrow **** \downarrow \downarrow \rangle$ and $|\psi_0 \rangle = |\downarrow \downarrow * \uparrow \downarrow * \downarrow \uparrow \downarrow \downarrow \downarrow \uparrow \rangle$ (where $*$ denotes a highly entangled spin).

We see that these examples present a behaviour with a survival probability quickly falling to zero as a scattering state but with strong resurgences of non-zero probabilities (lower than one) as a chaotic state. This another example of the hybrid nature of the chimera state.\\

\section{Conclusion}
The system defined by the Hamiltonian (\ref{ham}) exhibits hybrid behaviors between a decoherent and a coherent system. The chimera states of spin chains, presenting both highly entangled regions and totally not entangled regions. In contrast with the classical system (\ref{clchimere}), the quantum chimera states are stable (infinite life duration) in spite of the relatively small number of subsystems, because they are eigenstates. This is due to the fact that our quantum subsystems are spins and not oscillators, the ferromagnetic interactions present in the Hamiltonian (\ref{ham}) stabilise the chain. Moreover the number of subsystems in our quantum model (\ref{ham}) is lower than the minimum of subsystems needed in the classical one (\ref{clchimere}). This can be also a consequence of the difference between the two kinds of subsystems (oscillators and spins). But this could be also a consequence of the non-local nature of the quantum mechanics which could be make more efficient the effects of the long range couplings.\\
The quantum chimera states could be very interesting for quantum information protocoles. We could imagine transports of information using the couplings of the chain from a region to another one with manipulations taking advantage of the radical difference of the entanglement amplitudes. The model presented in the present paper has been constructed to be very close to the classical model (\ref{clchimere}) (i.e. with the same structure of couplings) and constitutes only a \textit{toy model}. It could present some unnecessary complexities and it would be interesting to study what are the necessary ingredients to involve quantum chimera states. It seems that the non hermiticity is needed. Futur works must be dedicated to find a more realistic model exhibiting quantum chimera states, but we can imagine experimental realizations of some quantum chimera models with trapped ions and cavity QED, since these situations can be modelled by non-hermitian effective Hamiltonians as in \cite{Lee1,Lee2}. Unfortunately, in these cases the effective Hamiltonian is valid only between two quantum jumps. In this context, it could be difficult to enlighten the dynamical behaviour of the chimera states (as the survival probabilities) since the characteristic time between two quantum jumps can be shorter than the time necessary to see a resurgence of the survival probability. Future works need to adapt the present \textit{toy model} to be closer to a concrete experimental situation where the possible chaotic dynamical behaviours will be more obvious. Nevertheless, fortunately the fundamental property of the quantum chimera states (one part of the chain is highly entangled whereas the other one is completely disentangled) is a stationary property. It can be highlighted or used in a rapid quantum protocol between two quantum jumps in possible experimental situations.


\begin{thebibliography}{0}
\bibitem{modelmixte} We have also considered a long range hermitian closed spin chain with Hamiltonian similar to equation (\ref{ham}) but with the matrices $\sigma_+$ substituted by $\sigma_x$.
\end{thebibliography}
\end{document}